\documentclass[twocolumn,letterpaper,aps,prl,superscriptaddress,showpacs,floatfix]{revtex4}

\usepackage{graphicx}	% Include figure filed

\usepackage{xspace}	% Include xspace
\usepackage{changebar}

\newcommand{\dphi}{\mbox{$\Delta\phi$}\xspace}
\newcommand{\deta}{\mbox{$\Delta\eta$}\xspace}
\newcommand{\pt}{\mbox{$p_T$}\xspace}

\newcommand{\sqsn}{\mbox{$\sqrt{s_{_{NN}}}$}\xspace}

\begin{document}

\title{Quadrupole Anisotropy in Dihadron Azimuthal
  Correlations in Central $d$$+$Au Collisions at $\sqrt{s_{_{NN}}}$=200~GeV 
}

\newcommand{\abilene}{Abilene Christian University, Abilene, Texas 79699, USA}
\newcommand{\augie}{Department of Physics, Augustana College, Sioux Falls, South Dakota 57197, USA}
\newcommand{\banaras}{Department of Physics, Banaras Hindu University, Varanasi 221005, India}
\newcommand{\barc}{Bhabha Atomic Research Centre, Bombay 400 085, India}
\newcommand{\baruch}{Baruch College, City University of New York, New York, New York, 10010 USA}
\newcommand{\bnlcoll}{Collider-Accelerator Department, Brookhaven National Laboratory, Upton, New York 11973-5000, USA}
\newcommand{\bnlphys}{Physics Department, Brookhaven National Laboratory, Upton, New York 11973-5000, USA}
\newcommand{\caucr}{University of California - Riverside, Riverside, California 92521, USA}
\newcommand{\charlesczech}{Charles University, Ovocn\'{y} trh 5, Praha 1, 116 36, Prague, Czech Republic}
\newcommand{\chonbuk}{Chonbuk National University, Jeonju, 561-756, Korea}
\newcommand{\ciae}{Science and Technology on Nuclear Data Laboratory, China Institute of Atomic Energy, Beijing 102413, P.~R.~China}
\newcommand{\cns}{Center for Nuclear Study, Graduate School of Science, University of Tokyo, 7-3-1 Hongo, Bunkyo, Tokyo 113-0033, Japan}
\newcommand{\colorado}{University of Colorado, Boulder, Colorado 80309, USA}
\newcommand{\columbia}{Columbia University, New York, New York 10027 and Nevis Laboratories, Irvington, New York 10533, USA}
\newcommand{\czechtech}{Czech Technical University, Zikova 4, 166 36 Prague 6, Czech Republic}
\newcommand{\dapnia}{Dapnia, CEA Saclay, F-91191, Gif-sur-Yvette, France}
\newcommand{\elte}{ELTE, E{\"o}tv{\"o}s Lor{\'a}nd University, H - 1117 Budapest, P{\'a}zm{\'a}ny P. s. 1/A, Hungary}
\newcommand{\ewha}{Ewha Womans University, Seoul 120-750, Korea}
\newcommand{\fit}{Florida Institute of Technology, Melbourne, Florida 32901, USA}
\newcommand{\fsu}{Florida State University, Tallahassee, Florida 32306, USA}
\newcommand{\gsu}{Georgia State University, Atlanta, Georgia 30303, USA}
\newcommand{\hiroshima}{Hiroshima University, Kagamiyama, Higashi-Hiroshima 739-8526, Japan}
\newcommand{\ihepprot}{IHEP Protvino, State Research Center of Russian Federation, Institute for High Energy Physics, Protvino, 142281, Russia}
\newcommand{\illuiuc}{University of Illinois at Urbana-Champaign, Urbana, Illinois 61801, USA}
\newcommand{\inrras}{Institute for Nuclear Research of the Russian Academy of Sciences, prospekt 60-letiya Oktyabrya 7a, Moscow 117312, Russia}
\newcommand{\instpasczech}{Institute of Physics, Academy of Sciences of the Czech Republic, Na Slovance 2, 182 21 Prague 8, Czech Republic}
\newcommand{\isu}{Iowa State University, Ames, Iowa 50011, USA}
\newcommand{\jaea}{Advanced Science Research Center, Japan Atomic Energy Agency, 2-4 Shirakata Shirane, Tokai-mura, Naka-gun, Ibaraki-ken 319-1195, Japan}
\newcommand{\jyvaskyla}{Helsinki Institute of Physics and University of Jyv{\"a}skyl{\"a}, P.O.Box 35, FI-40014 Jyv{\"a}skyl{\"a}, Finland}
\newcommand{\kek}{KEK, High Energy Accelerator Research Organization, Tsukuba, Ibaraki 305-0801, Japan}
\newcommand{\korea}{Korea University, Seoul, 136-701, Korea}
\newcommand{\kurchatov}{Russian Research Center ``Kurchatov Institute", Moscow, 123098 Russia}
\newcommand{\kyoto}{Kyoto University, Kyoto 606-8502, Japan}
\newcommand{\labllr}{Laboratoire Leprince-Ringuet, Ecole Polytechnique, CNRS-IN2P3, Route de Saclay, F-91128, Palaiseau, France}
\newcommand{\lahorelums}{Physics Department, Lahore University of Management Sciences, Lahore, Pakistan}
\newcommand{\lawllnl}{Lawrence Livermore National Laboratory, Livermore, California 94550, USA}
\newcommand{\losalamos}{Los Alamos National Laboratory, Los Alamos, New Mexico 87545, USA}
\newcommand{\lpc}{LPC, Universit{\'e} Blaise Pascal, CNRS-IN2P3, Clermont-Fd, 63177 Aubiere Cedex, France}
\newcommand{\lund}{Department of Physics, Lund University, Box 118, SE-221 00 Lund, Sweden}
\newcommand{\maryland}{University of Maryland, College Park, Maryland 20742, USA}
\newcommand{\mass}{Department of Physics, University of Massachusetts, Amherst, Massachusetts 01003-9337, USA }
\newcommand{\michigan}{Department of Physics, University of Michigan, Ann Arbor, Michigan 48109-1040, USA}
\newcommand{\muenster}{Institut fur Kernphysik, University of Muenster, D-48149 Muenster, Germany}
\newcommand{\muhlenberg}{Muhlenberg College, Allentown, Pennsylvania 18104-5586, USA}
\newcommand{\myongji}{Myongji University, Yongin, Kyonggido 449-728, Korea}
\newcommand{\nagasaki}{Nagasaki Institute of Applied Science, Nagasaki-shi, Nagasaki 851-0193, Japan}
\newcommand{\newmex}{University of New Mexico, Albuquerque, New Mexico 87131, USA }
\newcommand{\nmsu}{New Mexico State University, Las Cruces, New Mexico 88003, USA}
\newcommand{\ohio}{Department of Physics and Astronomy, Ohio University, Athens, Ohio 45701, USA}
\newcommand{\ornl}{Oak Ridge National Laboratory, Oak Ridge, Tennessee 37831, USA}
\newcommand{\orsay}{IPN-Orsay, Universite Paris Sud, CNRS-IN2P3, BP1, F-91406, Orsay, France}
\newcommand{\peking}{Peking University, Beijing 100871, P.~R.~China}
\newcommand{\pnpi}{PNPI, Petersburg Nuclear Physics Institute, Gatchina, Leningrad region, 188300, Russia}
\newcommand{\riken}{RIKEN Nishina Center for Accelerator-Based Science, Wako, Saitama 351-0198, Japan}
\newcommand{\rikjrbrc}{RIKEN BNL Research Center, Brookhaven National Laboratory, Upton, New York 11973-5000, USA}
\newcommand{\rikkyo}{Physics Department, Rikkyo University, 3-34-1 Nishi-Ikebukuro, Toshima, Tokyo 171-8501, Japan}
\newcommand{\saopaulo}{Universidade de S{\~a}o Paulo, Instituto de F\'{\i}sica, Caixa Postal 66318, S{\~a}o Paulo CEP05315-970, Brazil}
\newcommand{\stonybrkc}{Chemistry Department, Stony Brook University, SUNY, Stony Brook, New York 11794-3400, USA}
\newcommand{\stonycrkp}{Department of Physics and Astronomy, Stony Brook University, SUNY, Stony Brook, New York 11794-3400, USA}
\newcommand{\tenn}{University of Tennessee, Knoxville, Tennessee 37996, USA}
\newcommand{\titech}{Department of Physics, Tokyo Institute of Technology, Oh-okayama, Meguro, Tokyo 152-8551, Japan}
\newcommand{\tsukuba}{Institute of Physics, University of Tsukuba, Tsukuba, Ibaraki 305, Japan}
\newcommand{\vandy}{Vanderbilt University, Nashville, Tennessee 37235, USA}
\newcommand{\waseda}{Waseda University, Advanced Research Institute for Science and Engineering, 17 Kikui-cho, Shinjuku-ku, Tokyo 162-0044, Japan}
\newcommand{\weizmann}{Weizmann Institute, Rehovot 76100, Israel}
\newcommand{\wigner}{Institute for Particle and Nuclear Physics, Wigner Research Centre for Physics, Hungarian Academy of Sciences (Wigner RCP, RMKI) H-1525 Budapest 114, POBox 49, Budapest, Hungary}
\newcommand{\yonsei}{Yonsei University, IPAP, Seoul 120-749, Korea}
\affiliation{\abilene}
\affiliation{\augie}
\affiliation{\banaras}
\affiliation{\barc}
\affiliation{\baruch}
\affiliation{\bnlcoll}
\affiliation{\bnlphys}
\affiliation{\caucr}
\affiliation{\charlesczech}
\affiliation{\chonbuk}
\affiliation{\ciae}
\affiliation{\cns}
\affiliation{\colorado}
\affiliation{\columbia}
\affiliation{\czechtech}
\affiliation{\dapnia}
\affiliation{\elte}
\affiliation{\ewha}
\affiliation{\fit}
\affiliation{\fsu}
\affiliation{\gsu}
\affiliation{\hiroshima}
\affiliation{\ihepprot}
\affiliation{\illuiuc}
\affiliation{\inrras}
\affiliation{\instpasczech}
\affiliation{\isu}
\affiliation{\jaea}
\affiliation{\jyvaskyla}
\affiliation{\kek}
\affiliation{\korea}
\affiliation{\kurchatov}
\affiliation{\kyoto}
\affiliation{\labllr}
\affiliation{\lahorelums}
\affiliation{\lawllnl}
\affiliation{\losalamos}
\affiliation{\lpc}
\affiliation{\lund}
\affiliation{\maryland}
\affiliation{\mass}
\affiliation{\michigan}
\affiliation{\muenster}
\affiliation{\muhlenberg}
\affiliation{\myongji}
\affiliation{\nagasaki}
\affiliation{\newmex}
\affiliation{\nmsu}
\affiliation{\ohio}
\affiliation{\ornl}
\affiliation{\orsay}
\affiliation{\peking}
\affiliation{\pnpi}
\affiliation{\riken}
\affiliation{\rikjrbrc}
\affiliation{\rikkyo}
\affiliation{\saopaulo}
\affiliation{\stonybrkc}
\affiliation{\stonycrkp}
\affiliation{\tenn}
\affiliation{\titech}
\affiliation{\tsukuba}
\affiliation{\vandy}
\affiliation{\waseda}
\affiliation{\weizmann}
\affiliation{\wigner}
\affiliation{\yonsei}
\author{A.~Adare} \affiliation{\colorado}
\author{C.~Aidala} \affiliation{\mass} \affiliation{\michigan}
\author{N.N.~Ajitanand} \affiliation{\stonybrkc}
\author{Y.~Akiba} \affiliation{\riken} \affiliation{\rikjrbrc}
\author{H.~Al-Bataineh} \affiliation{\nmsu}
\author{J.~Alexander} \affiliation{\stonybrkc}
\author{A.~Angerami} \affiliation{\columbia}
\author{K.~Aoki} \affiliation{\kyoto} \affiliation{\riken}
\author{N.~Apadula} \affiliation{\stonycrkp}
\author{Y.~Aramaki} \affiliation{\cns} \affiliation{\riken}
% ...
\collaboration{PHENIX Collaboration} \noaffiliation

\date{\today}

\begin{abstract}

The PHENIX collaboration at the Relativistic Heavy Ion Collider 
(RHIC) reports measurements of azimuthal dihadron correlations near 
midrapidity in $d$$+$Au collisions at $\sqrt{s_{_{NN}}}$=200~GeV. 
These measurements complement recent analyses by experiments at the 
Large Hadron Collider (LHC) involving central $p$$+$Pb collisions at 
$\sqrt{s_{_{NN}}}$=5.02~TeV, which have indicated strong anisotropic 
long-range correlations in angular distributions of hadron pairs. 
The origin of these anisotropies is currently unknown. Various 
competing explanations include parton saturation and hydrodynamic 
flow. We observe qualitatively similar, but larger, anisotropies in 
$d$$+$Au collisions compared to those seen in $p$$+$Pb collisions at 
the LHC. The larger extracted $v_2$ values in $d$$+$Au collisions at RHIC 
are consistent with expectations from hydrodynamic calculations owing to 
the larger expected initial-state eccentricity compared with that 
from $p$$+$Pb collisions. 
When both are divided by an estimate of the initial-state eccentricity
the scaled anisotropies follow a common trend with multiplicity
that may extend to heavy ion data at RHIC and the LHC, where
the anisotropies are widely thought to arise from hydrodynamic flow.

\end{abstract}

	
\maketitle

% PRL length command
%\textbf{*** page break for PRL word count ***} \clearpage

Proton- and deuteron-nucleus collisions at relativistic energies are 
studied to provide baseline measurements for heavy-ion collision
measurements.  
In $p$($d$)+A collisions, initial-state nuclear effects are present; 
however, the formation of hot quark-gluon matter as created in heavy 
ion collisions is not commonly expected.  Recently there has been 
significant interest in the physics of high-multiplicity events in 
small collision systems, motivated by the observation of a small 
azimuthal angle ($\Delta\phi$) large pseudorapidity ($\Delta\eta$) 
correlation of primarily low $p_T$ particles in very high 
multiplicity $p$$+$$p$ collisions at 
7~TeV~\cite{Khachatryan:2010gv}. The correlation resembles the 
``near-side ridge" observed in 
Au$+$Au~\cite{Abelev:2009af,Alver:2009id}. The initial $p$$+$$p$ 
result sparked considerable theoretical 
interest~\cite{Dumitru:2010iy,Werner:2010ss,Dusling:2012iga}. 
Recently, a similar effect was observed in $p$$+$Pb collisions at 
\sqsn~=~5.02~TeV~\cite{CMS:2012qk}. Subsequent work from 
ALICE~\cite{Abelev:2012cya} and ATLAS~\cite{Aad:2012} removed 
centrality independent correlations (largely from jet fragmentation) 
by looking at the difference in correlations between central and 
peripheral events and has additionally uncovered similar long-range 
$\Delta\eta$ correlations at $\dphi~\approx~\pi$ beyond those 
expected from fragmentation of recoiling jets. The effect appears as 
a longitudinally extended azimuthal modulation with a predominantly 
quadrupole component [i.e. $\cos(2\Delta\phi$) and bears a 
qualitative resemblance in both magnitude and $p_T$ dependence to 
elliptic flow measurements in heavy ion collisions, where the large 
quadrupole modulation is understood to be caused by the 
initial-state spatial anisotropy followed by a nearly inviscid 
hydrodynamic expansion~\cite{Huovinen:2006jp}. A variety of physical 
mechanisms have been invoked to explain the observed anisotropies in 
$p$$+$Pb including gluon 
saturation~\cite{McLerran:2008pe,Dusling:2012iga,Dusling:2012wy,Dusling:2013oia}, 
hydrodynamics~\cite{Werner:2010ss,Bozek:2011if,Shuryak:2013ke}, 
multiparton interactions~\cite{Ryskin:2011mk}, and final-state 
expansion effects~\cite{Avsar:2010rf}.

Previous analyses involving two-particle correlations from $d$$+$Au
collisions at RHIC have not indicated any long-range features at small
$\dphi$~\cite{Abelev:2009af,Adare:2011sc,Adler:2005ad,Adler:2006hi}. 
However, these measurements involved
$p_T$ selections that emphasize jetlike correlations, rather
than the underlying event.
Also, Refs.~\cite{Adler:2005ad,Adler:2006hi} were based on $d$$+$Au collisions
recorded in 2003 with a small data sample, which limited the
statistical significance of the results. 

We present here the first analysis of very central $d$$+$Au events 
to measure hadron correlations between midrapidity particles at 
\sqsn~=~200~GeV. The center of mass energy per nucleon is a factor 
of 25 lower than at the LHC. Another potentially key difference is 
the use of a deuteron as the projectile nucleus rather than a 
proton. In Ref.~\cite{Bozek:2011if}, within the context of a 
Monte Carlo-Glauber (MC-Glauber) model, the calculated initial spatial 
eccentricity of the participating nucleons, $\varepsilon_{2}$, for 
central (large number of participants) $d$+Pb is more than a 
factor of 2 larger than in central $p$$+$Pb collisions at LHC 
energies. We find the initial spatial eccentricity $\varepsilon_2$ 
from the MC-Glauber model~\cite{Alver:2008aq} for $d$$+$Au at RHIC 
energies to be similar to the $d$+Pb calculations at LHC energies.

The results presented here are based on 1.56 billion minimum-bias
$d$$+$Au collisions at \sqsn~=~200~GeV recorded with the PHENIX~\cite{Adcox:2003zm}
detector in 2008. The event centrality in $d$$+$Au is determined from 
the integrated charge  measured by a
beam-beam counter facing the incoming Au
nucleus~\cite{Adare:2012qf}. Here we isolate a more central sample
than previously analyzed, to compare more closely to the LHC results.
We use central and peripheral event samples comprising the top 5\%
and 50\%--88\% of the total charge distributions, respectively.

This analysis considers charged hadrons measured within the two PHENIX
central arm spectrometers. Each arm covers nominally $\pi/2$ in azimuth and has
a pseudorapidity acceptance of $|\eta|<0.35$. Charged tracks are
reconstructed using drift chambers with a hit association
requirement in two layers of multiwire proportional chambers with pad
readout; the momentum resolution is 0.7\%
$\oplus$ 1.1\%$p$ (GeV/$c$). 
Electrons are rejected with a veto in the ring-imaging 
\v{C}erenkov counters.

All pairs satisfying the tracking cuts within an event
are measured. 
The yield of pairs satisfying tracking and particle identification cuts is
corrected for azimuthal acceptance through the use of mixed-event
distributions.  The conditional
yield of pairs is determined by
$  \frac{1}{N^{t}}\frac{dN^{\rm pairs}}{d\Delta\phi} \propto
   \frac{ dN^{\rm pairs}_{\rm same}/d\Delta\phi }{
    dN^{\rm pairs}_{\rm mix}/d\Delta\phi }$
where $N^{t}$ is the number of {\it trigger} hadrons (trigger
hadrons are those having the momenta required to begin the
search for a pair of hadrons)
and $N^{\rm pairs}_{\rm same}$ ($N^{\rm pairs}_{\rm mix}$) is the number of pairs from
the same (mixed) events. Mixed pairs are constructed with
particles from different events within the same 5\% centrality class and with
event vertices within 5~cm of each other.  Because the focus of this
analysis is on the shape of the distributions, no correction is
applied for the track reconstruction efficiency, which has a
negligible dependence on centrality for $d$$+$Au track multiplicities.

To make direct comparisons between our measurements and recent ATLAS 
$p$$+$Pb results~\cite{Aad:2012}, we follow a similar analysis 
procedure.  Charged hadrons with $0.5<p_T<3.5$~GeV/$c$ are used.  
For this analysis, each pair includes at least one particle at low $p_T$ 
($0.5<p_{T}<0.75$~GeV/$c$), which enhances the sensitivity to the 
nonjet phenomena.  To minimize the contribution from small-angle 
correlations arising from resonances, Bose-Einstein correlations, 
and jet fragmentation, pairs are restricted to pseudorapidity 
separations of $0.48<|\Delta\eta|<0.7$.  This $\Delta\eta$ gap is 
chosen to be as large as possible within the tracking acceptance, 
while still preserving an adequate statistical sample size. Unlike 
measurements at the LHC, this method is not sensitive to the 
pseudorapidity extent of the correlations.

%%%%%%%%%%%%%%%%%%%%%%%%%%%%%%%%%%%%%%%%% Fig_1
\begin{figure}[tb]
  \includegraphics[width=1.0\linewidth]{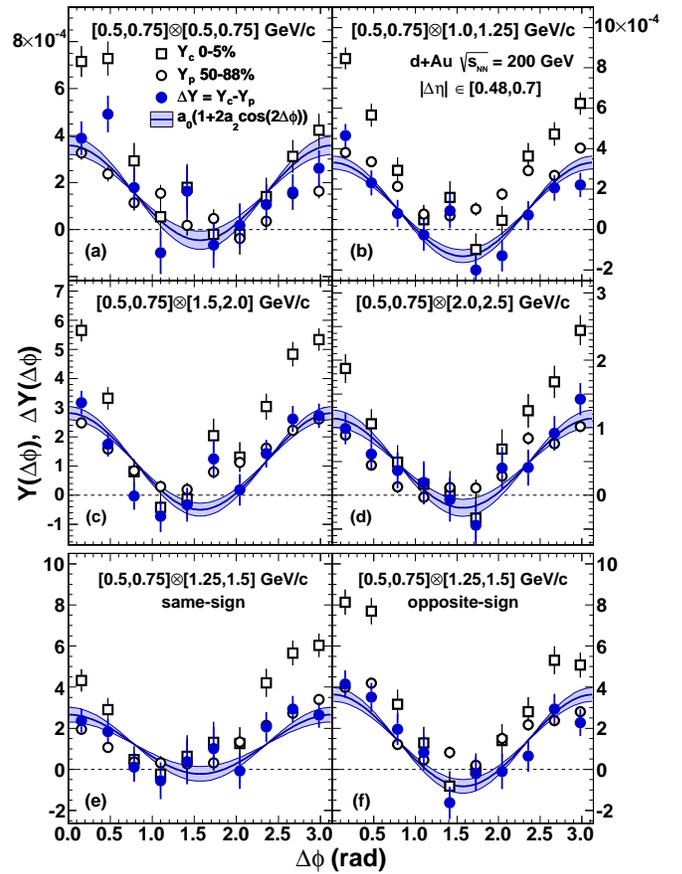}
  \caption{\label{fig1U2} 
Azimuthal conditional yields,
    $Y\left(\Delta\phi\right)$, for 
(open [black] squares) 0\%--5\% most central and 
(open [black] circles) peripheral (50\%--88\% least central) 
collisions with a minimum \deta separation of 0.48 units.  
Difference $\Delta Y (\dphi)$ (filled [blue] circles), which is 
([blue] curve) fit to $a_0 + 2a_2 \cos(2\Delta\phi)$, where 
$a_0$ and $a_2$ are computed directly from the data. (shaded 
[blue] band) Statistical uncertainty on $a_2$.  The bottom left 
(right) panel shows the same quantity for same-sign 
(opposite-sign) pairs.}

\end{figure}

The conditional yield owing to azimuthally uncorrelated 
background is estimated by means of the zero-yield-at-minimum (ZYAM)
procedure~\cite{Ajitanand:2005jj}. This background contribution is
obtained for both the central  and peripheral samples by
performing fits to the conditional yields using a functional form
composed of a constant pedestal and two Gaussian peaks, centered at
$\dphi=0$ and $\pi$. The minimum of this function,
$b_{\rm ZYAM}$, is subtracted from the conditional yields,
and the result is:
$Y(\Delta\phi) \equiv \frac{1}{N^t}\frac{dN^{\rm pairs}}{d\Delta\phi} - b_{\rm ZYAM}$
 The conditional yields
$Y_c(\dphi)$ and $Y_p(\dphi)$ (central and peripheral events,
respectively) are shown in Fig.~\ref{fig1U2}, along with their
difference $\Delta Y (\dphi) \equiv Y_c(\dphi) - Y_p(\dphi)$.
As in Ref.~\cite{Aad:2012},
this subtraction removes any centrality independent correlations, 
such as effects from unmodified jet fragmentation, resonances and HBT.
In the absence of any centrality dependence, $Y_c(\Delta\phi)$
and $Y_p(\Delta\phi)$ should be identical.  
It is notable that any signal in the peripheral events is subtracted from the central
events.  
We see
that $Y_c(\Delta\phi)$ is significantly larger than $Y_p(\Delta\phi)$
for $\Delta\phi$ near 0 and $\pi$.

%%%%%%%%%%%%%%%%%%%%%%%%%%%%%%%%%%%%%%%%% Fig_2
\begin{figure}[tb]
  \includegraphics[width=1.0\linewidth]{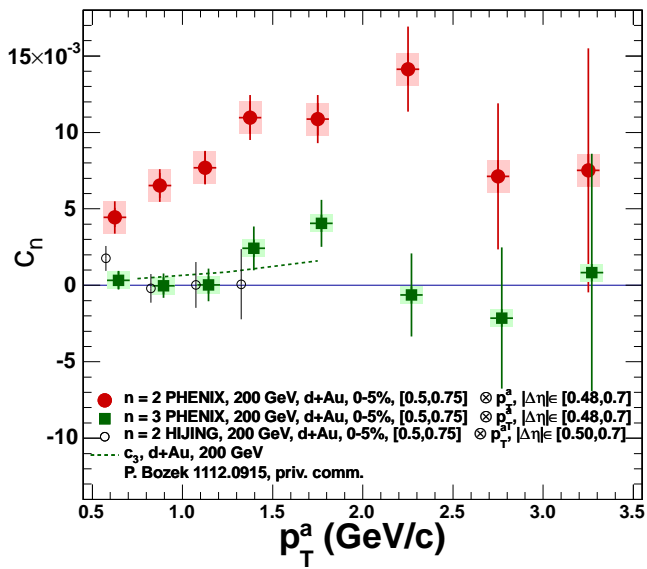}
  \caption{\label{fig3} 
The $n$th-order pair anisotropy, $c_{n}$, of the central collision 
excess as a function of associated particle $p^{a}_{T}$.
$c_2$ (filled [red] circles) and 
$c_3$ (filled [green] squares) are for 
$0.5<p^{t}_{T}<0.75$~GeV/$c$, $0.48<|\Delta\eta|<0.7$ 
$c_2$ as extracted from $d+$Au {\sc hijing} events
using the same procedure as in the data is 
also
shown (open circles).
$c_3$ as expected for our $p_{T}$ selections from
Ref. [31] is shown as a dashed line.}
\end{figure}

We find that the
difference with centrality is well described by the
symmetric form:
$\Delta Y (\dphi) \approx  a_{0} + 2a_{2} \cos\left( 2\dphi \right)$
as demonstrated in Fig.~\ref{fig1U2}. The  coefficients $a_{n}$
and their statistical uncertainties are computed from the $\Delta
Y(\dphi)$ distributions as:
$a_{n} = \langle \Delta Y(\dphi) \cos(n\dphi) \rangle$.
The $\cos(2\dphi)$ modulation appears as the dominant component of the
anisotropy for all $p_T$ combinations.

%%%%%%%%%%%%%%%%%%%%%%%%%%%%%%%%%%%%%%%%% xxxx combined with old Fig. 1
%\begin{figure}[tb]
%  \includegraphics[width=1.0\linewidth]{fig2}
%  \caption{\label{fig2} (color online)
%Sample comparison of $Y\left(\Delta\phi\right)$ and
%$\Delta Y\left(\Delta\phi\right)$ for same and 
%oppositely charged pairs for $1.25<p^{a}_{T}<1.5$
%GeV/$c$ and $0.48<|\Delta\eta|<0.7$.
%The symbols, curve, and shaded band are as 
%described in the Fig.~\protect\ref{fig1U2} caption. }
%\end{figure}

To quantify the relative amplitude of the azimuthal modulation, 
we define
$c_{n} \equiv a_{n} / \left( b^{c}_{\rm ZYAM} + a_{0} \right)$,
where $b^c_{\rm ZYAM}$ is $b_{\rm ZYAM}$ in central events.
$c_2$ and $c_3$ are shown as a function of associated $p_{T}$ in
Fig.~\ref{fig3} for central (0\%--5\%) collisions.

The dominant source of systematic uncertainty results from the 
inability to completely exclude the near-side jet peak in this 
analysis.  The PHENIX central arm spectrometers lack sufficient 
$|\deta|$ acceptance to completely exclude the near-side jet peak. 
To assess the systematic influence of any residual unmodified jet 
correlations, we analyzed charge-selected correlations. 
Charge ordering is a known feature of jet fragmentation, which 
leads to enhancement of the jet correlation in opposite-sign 
pairs, and suppression in like-sign pairs, in the near side peak 
(e.g. Ref.~\cite{Adler:2002tq}). A representative $p_T$ selection 
of $Y(\dphi)$ and $\Delta Y(\dphi)$ distributions for like-sign 
and opposite-sign pairs are shown in Fig.~\ref{fig1U2} (bottom 
panels). Both $\Delta Y(\dphi)$ distributions exhibit a 
significant $\cos(2\Delta\phi)$ modulation. The magnitude of the 
modulation at $\Delta\phi=$~0 is larger in the opposite-sign case. 
The root-mean-squared variation of the same-sign and opposite-sign 
$c_n$ measurements relative to the combined value is included in 
the systematic uncertainties.  This reflects the influence of 
possible remaining jet correlations and is applied symmetrically, 
because the influence of the jet contribution is not known. As an 
additional test, the minimum $\deta$ was varied from the nominal 
value of 0.48 to 0.36 (where sensitivity to jet contributions is 
enhanced) and 0.60 (where it is reduced). The 0.36 selection has 
some $\dphi$ asymmetry in $\Delta Y(\dphi)$; the 0.60 selection 
does not.  In both cases the extracted $c_2$ values are consistent 
with the central $\Delta\eta$ selection.  To assess the dependence 
of the results on our selection of peripheral events, we have 
extracted c2 values using 60\%--88% and 70\%--88\% central events 
as alternate peripheral samples. No significant change was found 
in the $c_2$ values from the default peripheral subtraction.  
This is potentially different from the implications of 
Ref.~\cite{Adams:2004bi} where a difference in low $p_T$ hadron 
correlations between 40\%--100\% $d$$+$Au and $p$$+$$p$ collisions 
is observed. We observe a similar magnitude signal in both 
0\%--5\% and 0\%--20\% central events.  Other sources of 
uncertainty, such as occupancy and acceptance corrections, were 
found to have negligible effect on these results.

%The ATLAS $c_2$ results~\cite{Aad:2012} have a qualitatively 
%similar $p_T^a$ dependence, but with a significantly smaller magnitude.
%However, it must be noted that the $c_2$ values
%from PHENIX and ATLAS are not directly comparable since $c_2$ is a
%function of the $p_T$ of both particles and the trigger particle $p_T$
%range is not identical in the two analyses. ATLAS has also used a much
%larger $\Delta\eta$ separation between the particles.

In $p$$+$Pb collisions at the LHC the signal is seen in long-range $\Delta\eta$ correlations.
In this analysis, signal is measured at midrapidity, but it is natural to ask if
previous PHENIX rapidity separated correlation
measurements~\cite{Adare:2011sc} would have been sensitive to a signal
of this magnitude. The maximum $c_2$ observed here
is approximately a 1\% modulation about the background level.
Overlaying a modulation of this size on the conditional yields shown
in Fig. 1 of Ref.~\cite{Adare:2011sc} shows that the modulation on the
near side is small compared with the statistical uncertainties.
%In the current analysis, both particles are near midrapidity, while
% Ref.~\cite{Adare:2011sc} includes one of the particles
%very forward ($3.0<\eta<3.8$) in the $d$-going direction. 
With the current method
we cannot determine whether the signal observed here persists for $\eta>$3.

%In order to test if correlations between multiplicity
%selected centrality and jet energy or quark versus
%gluon jet contributions might mimic this effect, we have
To test effects of the centrality determination or
known jet modifications on this observable, we have
applied the identical analysis procedure (including the
centrality selection) to {\sc hijing}~\cite{Gyulassy:1994ew} (v1.383) $d+$Au 
events.
As shown in Fig.~\ref{fig3}, we find an average $c_2$
value 
of (7.5$\pm$5.5)$\times 10^{-4}$ for 0.5~$<p_T^a<$~1.5~GeV/$c$ with
no significant $p_T$ dependence.

The $c_3$ values, shown in Fig.~\ref{fig3}, are small relative to
$c_2$. Fitting the $c_3$ data to a constant yields (6$\pm$4)$\times
10^{-4}$ with a $\chi^2$ per degree of freedom of 8.4/7 (statistical 
uncertainties only); no significant $c_3$ is observed. 

%%%%%%%%%%%%%%%%%%%%%%%%%%%%%%%%%%%%%%%%% Fig_3
\begin{figure}[tb]
  \includegraphics[width=1.0\linewidth]{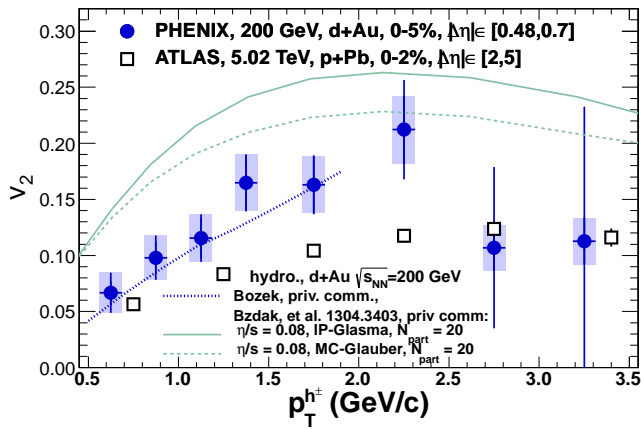}
  \caption{\label{fig4} 
Charged hadron second-order anisotropy, $v_2$,
as a function transverse momentum for 
(filled [blue] circles) PHENIX and 
(open [black] squares) ATLAS~\protect\cite{Aad:2012}.
Also shown are  hydrodynamic 
calculations from Bozek~[14,31] (dotted [blue] curve)
and Bzdak {\it et al.}~[32,39] 
for impact parameter glasma initial conditions (solid curve) 
and the MC-Glauber model initial conditions (dashed curve).}
\end{figure}

A measure of the single-particle anisotropy, $v_2$, can be obtained
under the assumption of factorization~\cite{Luzum:2010sp,Alver:2010dn,
Aamodt:2011by}:
$c_{2}\left( p^{t}_{T}, p^{a}_{T} \right) 
= v_2\left(p^{t}_{T}\right) v_2\left(p^{a}_{T}\right)$. 
We have varied $p^{t}_{T}$ and recomputed
$v_2\left(p_T\right)$ and find no significant deviation from the
factorization hypothesis.  The calculated single particle $v_2$ is
shown in Fig.~\ref{fig4}, and also compared with the
ATLAS~\cite{Aad:2012} results, revealing qualitatively similar
$\pt$ dependence with a significantly larger magnitude. We also compare the
$v_2$ results to a hydrodynamic
calculation~\cite{Bozek:2011if,Bozek:privatecomm} and find good
agreement between the data and the calculation.
The $v_2$ reported here is the excess $v_2$ beyond any which is present in peripheral $d$$+$Au
collisions.
While we cannot extract $v_3$ from the current data, Fig.~\ref{fig3} shows 
that the
measured $c_3$ values are in agreement with the values expected from
$v_3$ as a function of $p_T$ in the same model as the $v_2$ calculation~\cite{Bozek:privatecomm}.
The $v_2$ data are also in qualitative agreement with 
another hydrodynamic calculation~\cite{Bzdak:2013zma}
both with the MC-Glauber model and with 
impact-parameter glasma~\cite{Schenke:2012wb} 
initial conditions (note that these calculations are at a fixed $N_{\rm part}$, not
the exact centrality range as in the data).  These calculations have very different assumptions 
about the initial geometry and yet are all in qualitative agreement with the data.

\begin{figure}[t]
\includegraphics[width=1.0\linewidth]{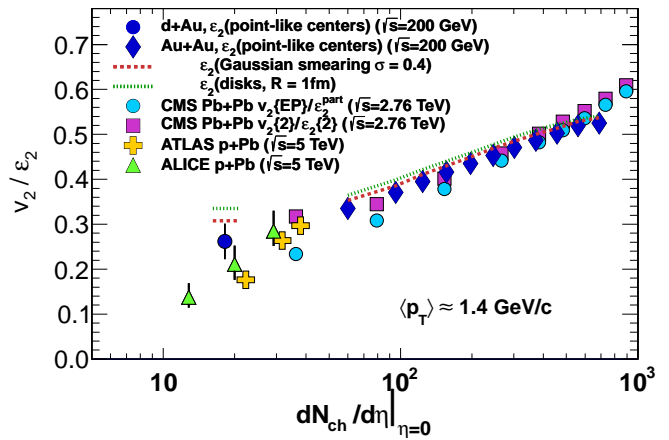}
\caption{\label{fig5} (color online)
The eccentricity-scaled anisotropy, $v_2/\varepsilon_2$,
vs charged-particle multiplicity ($dN_{\rm ch}/d\eta$) 
for $d$$+$A and $p$$+$Pb collisions~\protect\cite{Aad:2012,Abelev:2012cya}.   
Also shown are Au$+$Au data 
at \sqsn~=~200~GeV~\protect\cite{Adler:2004zn,Adare:2010ux,Lacey:2010fe}
and Pb+Pb data 
at \sqsn~=~2.76~TeV~\protect\cite{Chatrchyan:2012ta,Chatrchyan:2011pb}.  
The $v_2$ are for similar $p_T$ selections.  
The colored curves are for different nucleon representations in 
the $\varepsilon_2$ calculation in the MC-Glauber model.
The errors shown are statistical only and only shown on the $d$$+$Au point
with the point-like centers $\varepsilon_2$ for clarity.
Owing to the lack of available multiplicity data in p$+$Pb and 
$d$$+$Au collisions the $dN_{\rm ch}/d\eta$ values for those systems 
are calculated from {\sc hijing}~\cite{Gyulassy:1994ew}.  All $dN_{\rm 
ch}/d\eta$
values are in the center of mass system.
}
\end{figure}

To further investigate the origin of this effect, we plot in 
Fig.~\ref{fig5} the PHENIX results for both $d$$+$Au and Au$+$Au 
scaled by the eccentricity ($\varepsilon_2$), as calculated in a 
MC-Glauber model, as a function of the charged-particle 
multiplicity at midrapidity. Due to the lack of available 
multiplicity data for the $d$$+$Au centrality selection the $dN_{\rm 
ch}/d\eta$ value is calculated from HIJING~\cite{Gyulassy:1994ew}. 
The 0\%--5\% $d$$+$Au collisions at \sqsn~=~200~GeV have a $dN_{\rm 
ch}/{d\eta}$ similar to those of midcentral $p$$+$Pb collisions at 
the LHC, while the $\varepsilon_2$ values for $d$$+$Au collisions 
are about 50\% larger than those calculated for the midcentral 
$p$$+$Pb collisions. The key observation is that the ratio 
$v_2/\varepsilon_2$ is consistent between RHIC and the LHC, despite 
the factor of 25 difference in collision center of mass energy. A 
continuation of this trend is seen by also comparing to 
$v_2/\varepsilon_2$ as measured in 
Au$+$Au~\cite{Adler:2004zn,Adare:2010ux,Lacey:2010fe} and 
Pb+Pb~\cite{Chatrchyan:2012ta,Chatrchyan:2011pb} collisions. The 
$\varepsilon_2$ values calculated depend on the nucleon 
representation used in the MC-Glauber model.  In large systems 
this uncertainty is small, but in small systems, such as $d$$+$Au, 
this uncertainty becomes much more significant.  For illustration, 
$\varepsilon_2$ has been calculated using three 
different representations of the participating nucleons, pointlike 
centers, Gaussians with $\sigma$ = 0.4~fm, and uniform disks with 
R~=~1~fm for the PHENIX data.  The scaling feature is robust against 
these geometric variations, which leads to an approximately 30\% 
difference in the extracted $\varepsilon_2$ in $d$$+$Au collisions 
(other models, e.g. Ref.~\cite{Bzdak:2013zma}, could produce larger 
variations).

In summary, a two-particle anisotropy at midrapidity in the 5\% most
central $d$$+$Au collisions at \sqsn~=~200~GeV is observed. The excess
yield in central compared to peripheral events is well described by a
quadrupole shape. The signal is qualitatively similar,
but with a significantly larger amplitude than that observed
in long-range correlations in $p$$+$Pb collisions at much
higher energies.  While our
acceptance does not allow us to exclude the possibility of centrality
dependent modifications to the jet correlations, the subtraction of
the peripheral jetlike correlations has been checked both by varying
the $\Delta\eta$ cuts and exploiting the charge sign
dependence of jet-induced correlations. The observed results are in
agreement with a hydrodynamic calculation for $d$$+$Au collisions at
\sqsn~=~200~GeV.

%%%%%%%%%%%%%%%%%%%%%%%%%%%%%%%%%%%%%%%%% Fig_4

We find that scaling the results from RHIC and the LHC by the initial
second-order participant eccentricity from a MC-Glauber 
model~\cite{Bozek:2011if} may bring
the results to a common trend as a function of $dN_{\rm ch}/d\eta$.  
This may suggest that the phenomena observed here are 
sensitive to the initial state geometry and that the same 
underlying mechanism may be responsible in both $p$$+$Pb collisions at 
the LHC and $d$$+$Au collisions at RHIC.  It may also imply a 
relationship to the hydrodynamical understanding of $v_2$ in heavy 
ion collisions. The observation of $v_2$ at both RHIC 
and the LHC provides important new information.  Models intended to describe the
data must 
be capable of also explaining their persistence as the center of 
mass energy is varied by a factor of 25 from RHIC to the 
LHC.

% PRL length command
%\textbf{*** page break for PRL word count $<$3.5 pages $<$7 columns}
%\clearpage

%%%%%%%%%%%%%%%%%%%%%%%%% Acknowledgements

\begin{acknowledgments}

%\section{Acknowledgements}   % Run-8 long from for PRC, PLB, etc.

We thank the staff of the Collider-Accelerator and Physics
Departments at Brookhaven National Laboratory and the staff of
the other PHENIX participating institutions for their vital
contributions.  We acknowledge support from the 
Office of Nuclear Physics in the
Office of Science of the Department of Energy, the
National Science Foundation, Abilene Christian University
Research Council, Research Foundation of SUNY, and Dean of the
College of Arts and Sciences, Vanderbilt University (U.S.A),
Ministry of Education, Culture, Sports, Science, and Technology
and the Japan Society for the Promotion of Science (Japan),
Conselho Nacional de Desenvolvimento Cient\'{\i}fico e
Tecnol{\'o}gico and Funda\c c{\~a}o de Amparo {\`a} Pesquisa do
Estado de S{\~a}o Paulo (Brazil),
Natural Science Foundation of China (People's Republic of China),
Ministry of Education, Youth and Sports (Czech Republic),
Centre National de la Recherche Scientifique, Commissariat
{\`a} l'{\'E}nergie Atomique, and Institut National de Physique
Nucl{\'e}aire et de Physique des Particules (France),
Bundesministerium f\"ur Bildung und Forschung, Deutscher
Akademischer Austausch Dienst, and Alexander von Humboldt Stiftung (Germany),
Hungarian National Science Fund, OTKA (Hungary), 
Department of Atomic Energy and Department of Science and Technology (India), 
Israel Science Foundation (Israel), 
National Research Foundation and WCU program of the 
Ministry Education Science and Technology (Korea),
Ministry of Education and Science, Russian Academy of Sciences,
Federal Agency of Atomic Energy (Russia),
VR and Wallenberg Foundation (Sweden), 
the U.S. Civilian Research and Development Foundation for the
Independent States of the Former Soviet Union, 
the US-Hungarian Fulbright Foundation for Educational Exchange,
and the US-Israel Binational Science Foundation.

\end{acknowledgments}

%%%%%%%%%%%%%%%%%%%%%%%%%%% References

%\bibliography{ppg152x7}

\end{document}